\documentclass[prb,twocolumn,showpacs,amssymb]{revtex4}
\usepackage{graphicx}

\begin{document}

\title{Quantum Hall effects in layered disordered superconductors}

\author{V. Kagalovsky$^1$, B. Horovitz$^2$, Y. Avishai$^2$}
\affiliation{$^1$Negev Academic College of Engineering, Beer-Sheva 84100, Israel\\
$^2$Department of Physics and Ilze Katz center for
nanotechnology, Ben-Gurion University of the Negev,
Beer-Sheva 84105, Israel}

\begin{abstract}
Layered singlet paired superconductors with disorder and broken
time reversal symmetry are studied. The phase diagram demonstrates
charge-spin separation in transport. In terms of the average
intergrain transmission and the interlayer tunnelling we find
quantum Hall phases with spin Hall coefficients of
$\sigma_{xy}^{spin}=0,2$ separated by a spin metal phase. We
identify a spin metal-insulator localization exponent as well as a 
spin conductivity exponent of $\approx 0.9$. In presence of a Zeeman term an additional
$\sigma_{xy}^{spin}=1$ phase appears.

\end{abstract}


 \maketitle

The problem of quasiparticle transport and localization in
disordered superconductors is of considerable interest in view of
experimental activity on the high $T_c$ cuprates as well as
theoretical realization that disordered superconductors provide
new symmetry classes of random matrix theory\cite{altland}. Of
particular interest is class C for which the Hamiltonian breaks
time reversal symmetry  but spin rotation invariance remains
intact. Physically, it can be realized in materials consisting of
singlet superconductor grains in a magnetic field or else, by a
superconductor in the absence of magnetic field whose order
parameter breaks time reversal invariance, such as $d+id'$.  Class
C can therefore be realized by high $T_c$ compounds where d wave
pairing is well established. In fact, $d+id'$ pairing has been
suggested\cite{covington}, in particular in overdoped compounds or
as field induced pairing \cite{dagan}.

Transport properties of random superconductors are unusual since a
quasiparticle does not carry charge, being screened by the
condensate, while the singlet paired condensate does not transport
spin. Furthermore, the gapless nature of d wave pairing with low
lying quasiparticle excitations leads to a rich phase diagram in
2-dimensions (2D) with spin quantum Hall
phases\cite{su2,senthil1}, spin insulators and spin
metals\cite{senthil2,bundschuh};  a metallic phase was also found for triplet pairing \cite{read}.

The usual quantum Hall system in two dimensions (2D), as well as its extension to
3-dimensional (3D) layered system, have been studied by a network
model \cite{cc,cd} which consists of a lattice of nodes connected
by links. In 2D, the unidirectional propagation on links is described by random
phases, corresponding to a group U(1), while transfer at nodes is
controlled by a parameter which determines the critical point. The
transfer matrix of the network model can be efficiently evaluated
identifying the critical behavior, e.g. the localization exponent
is $\nu_{QH}\approx 2.5$. The 2D class C problem has recently been
studied by a network model \cite{su2} where propagation on links
of particle-hole spinors via the Bogoliubov-de Gennes Hamiltonian
is realized by random SU(2) matrices. The quantized spin Hall
conductance is
 shown to jump by two units at a critical point of a new universality
 class with a localization exponent $\nu_{2d}\approx 1.12$; an exact
 mapping on a classical percolation problem \cite{gruzberg,beamond} 
has found $\nu_{2d}=4/3$. The spin
 rotation invariance can be broken by having a different transmission
 for particles and holes\cite{su2}, e.g. a Zeeman term. The phase diagram has then 3 phases
 with quantum Hall values of 0, 1, 2, respectively and a localization
 exponent $\nu_{QH}$ of the usual U(1) theory.

 Experimental realization of this unusual spin transport depends
 on the ability to control deviations
 from the critical point. In the usual quantum Hall effect this is controlled
 by the position of the Fermi energy relative to that of an
 extended state in a
 Landau band. In a superconductor the particle-hole symmetry fixes
 the Fermi energy at the middle of the gap, and the relative position
 of states is not directly tuned by the overall density. It was in fact
 suggested that changing the strength of disorder can lead to quantum
 Hall transitions\cite{senthil1}, at least for weak breaking of time
 reversal symmetry.

An important insight into the nature of $d+id'$ superconductors
comes from studying their edge states\cite{covington,dagan,hg} which 
provide a realization of our network model and
 identify its parameters. In
 the $d$ wave case a prominent zero bias anomaly\cite{covington,dagan} has
 identified a surface state at zero energy. The $d+id'$ case allows
 current carrying chiral states that split the zero bias anomaly
 as seen in the overdoped compounds \cite{dagan}. The chirality of
 these edge states leads directly to quantized Hall
 conductance\cite{senthil1,hg}.  In contrast, 
 charge transport of superconducting grains
 is dominated by the randomness of the Josephson coupling between the grains; 
the phase correlation
 between grains is lost in 2D at a critical value of disorder, as
 shown in an XY model with random phase shifts\cite{carpentier}. 

 In the present work we  solve a network model for a
 layered 3-dimensional (3D) random superconductor. We find 3 phases which we
 identify as spin insulators with Hall coefficient $0$ and $2$,
 respectively, and a spin metal phase. We also identify the localization exponent at the insulator-metal transition, $\nu_{3d}\approx 0.9$. When the spin rotation
 symmetry is broken by a Zeeman term we find an additional phase with
 Hall coefficient of 1. The spin metal phase is a realization of
 a proposed phase with nonzero spin diffusion constant at zero
 temperature\cite{senthil2}. We identify the conductivity of the spin metal and find its critical exponent to be $\nu_{3d}$ as well.  Finally, we identify the physical parameter that
 controls criticality of spin transport, namely the average transmission of
 quasiparticles between grains. This then demonstrates
 spin-charge separation in the sense that their critical behavior
 relevant to transport is controlled by distinct parameters.

 The 3D network model  consists of layers of 2D lattices. Each 2D lattice has
 nodes, which are connected by links\cite{cc}. Quasiparticles
 propagate unidirectionally on links, hence their transfer matrix is
 equivalent to the evolution operator, which for a singlet
 superconductor is an $SU(2)$ matrix describing an Andreev process
 where particle and hole components mix\cite{su2},
 \begin{equation}
{\bf T}_1=\left( \begin{array}{cc}e^{i\delta_1}\sqrt{1-x} & -e^{i\delta_2}\sqrt{x}  \\
 e^{-i\delta_2}\sqrt{x} & e^{-i\delta_1}\sqrt{1-x}
\end{array}
\right),
\label{first}
\end{equation}
where $\delta_1$, $\delta_2$, $x$ are random, $0\leq \delta_1$, $\delta_2<2\pi$, $0\leq x<1$.

 The propagation between
 grains, i.e. at nodes of the network, is determined by the transmission
 between grains. At each node we have two incoming and two outgoing
 links with particles and holes separately scattered. The transmission
 probability at a node is parameterized in the form $T_0=[1+\exp (-\pi
 \epsilon)]^{-1}$, so that for $\epsilon=0$ the transmission equals
 reflection i.e. maximal mixing of all links.  In the
 following we allow also a Zeeman parameter $\Delta$ which breaks spin rotation
 invariance. The transfer matrix across a node is then
 \begin{equation}
{\bf T}_2=\left( \begin{array}{cc}\sqrt{1+e^{ -\pi
(\epsilon\pm\Delta /2)}} & e^{-\pi
(\epsilon /2\pm\Delta /4)}  \\
e^{-\pi (\epsilon /2\pm\Delta /4)}
& \sqrt{1+e^{-\pi (\epsilon\pm\Delta /2)}}
\end{array}
\right),
\label{second}
\end{equation}
where the $\pm$ sign corresponds to particle with spin-up or hole with
spin-down, respectively.

The final ingredient of the 3D network model 
are additional nodes\cite{cd} connecting neighboring layers. Placing 2D 
 layers one on top of the other allows for links belonging to
neighboring layers to form nodes and quasiparticles scatter. The matrix
describing this scattering is
\begin{equation}
{\bf T}_3=\left( \begin{array}{cc}\sqrt{1-t^2} & t  \\
-t & \sqrt{1-t^2}
\end{array}
\right).
\label{third}
\end{equation}
Consider a system of size $M \times M \times L$ where $M$ is the
number of links in one layer (with two channels per link) and $L
\to \infty$ is its length. For a given $M$ the eigenvalues of $T^{\dagger}T$, where $T$ is the full transfer matrix, behave as 
$e^{-2\lambda_nL}$, defining the Lyapunov exponents $\{\lambda_n\}$; the smallest positive one, $\lambda_1$, defines the localization length $\xi_M=1/\lambda_1$. 
The $M$ dependence of $\xi_{M}/M$
identifies the phases: (i) a decreasing ratio corresponds to localized
state, i.e. a spin insulator, (ii) a constant ratio corresponds to a critical
state, and (iii) an increasing ratio corresponds to a spin metal. 
The phase diagram for the class C network model (with
$\Delta=0$) is displayed in Fig. 1. Square boxes represent
critical $\epsilon_{cr}(t)$ lines. The particle-hole symmetry of the
superconductor ensures a degeneracy at the critical
point $\epsilon=t=0$, i.e. the Hall coefficient changes by two
units. Furthermore, in the clean limit the Hall conductance has
two units\cite{hg}; this corresponds to transmission $T_0=1$, i.e. $\epsilon>0$ large. Hence there are three distinct phases: Hall
insulator with Hall conductance $\sigma_{xy}=0$, spin metallic
phase, and a quantized spin Hall phase with $\sigma_{xy}=2$.

\begin{figure}
%[htb]
\begin{center}
\includegraphics[scale=0.7]{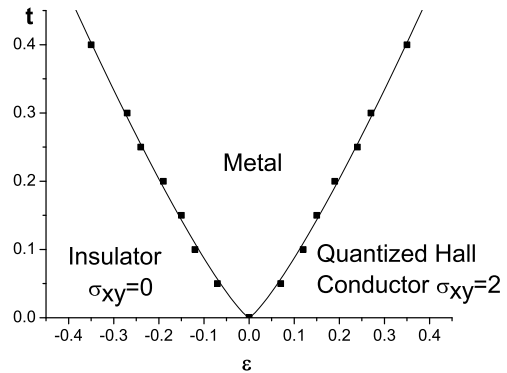}
\end{center}
\caption{Phase diagram of the 3D network model with
spin-rotational symmetry.}
\end{figure}

The width $W(t)$ in $\epsilon$ of the metallic region increases
with $t$, and is expected to behave as\cite{cd} $W(t)\sim
t^{1/\nu_{2d}}$ where $\nu_{2d}$ is the localization length
exponent in 2D. The argument is that for a 2D isolated layer the
mean level spacing is $\sim 1/\xi^{2}_{2d}$ with
$\xi_{2d}$ the 2D localization length. The states in each layer
are concentrated along a percolation cluster of length $\xi_{2d}$
and width of the edge state (a coherence length\cite{hg}) , i.e.
normalized as $\sim \xi^{-1/2}_{2d}$; the interlayer coupling is
then $\sim t/\xi_{2d}$. At the mobility edge the mean level spacing is
of the same order as the interlayer coupling according to the
Thouless criterion, hence $\epsilon \sim t^{1/\nu_{2d}}$. The
curve in Fig. 1. represents the least square fit for the data,
producing $W\sim t^{1/1.2}$, which is in good agreement with our
previous value \cite{su2} of $\nu_{2d}$.

The divergence of the localization length at  $\epsilon_{cr}(t)$
identifies the localization exponent of a
spin insulator-metal transition in 3D. We have evaluated the
critical exponent at the symmetric point
$t=1/\sqrt{2}$ (larger $t$ maps into a smaller t by rearranging layer indices)
 and found $\nu_{3d}=0.91$ on both the insulator and metal sides, as shown in Fig. 2; 
the same value is found for $t=0.1$.
 
\begin{figure}
%[htb]
\begin{center}
\includegraphics[scale=0.7]{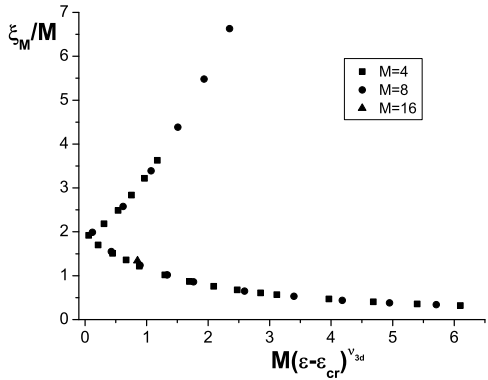}
\end{center}
\caption{Scaling of renormalized localization length for the 3D
network model with spin-rotational symmetry at $t=1/\sqrt{2}$
showing an exponent $\nu_{3d}=0.91$. The lower and upper branches correspond to the insulating $\epsilon>\epsilon_{cr}$ and metallic $\epsilon<\epsilon_{cr}$ phases, respectively. }
\end{figure}
 
 Fig. 2 shows that in the metallic phase $\xi_M/M$ increases 
 approximately linearly with $M$. It was proposed \cite{mackinnon}
 that this identifies the 3D conductivity as 
 $\sigma_{xx}\sim (\epsilon-\epsilon_{cr})^{\nu_{3d}}$. 
 This derivation needs to be revised since
 in the 3D limit the conductivity involves  many Lyapunov exponents $\lambda_n$. The multichannel conductance is given by  \cite{imry,pichard} $g=\sum_n[1+\cosh (\lambda_nL)]^{-1}$. For a few channels $M^2\ll L$ the lowest Lyapunov dominates, but in the 3D limit $M^2\gg L$ the number of terms $N_{eff}$ that contribute to $g$ is large. In fact, for many channels the rigidity in the spectrum of $T^{\dagger}T$ suppresses fluctuations in $\{\lambda_n\}$ and one expects \cite{pichard} $\lambda_n=n\lambda_1$. Hence $g\approx N_{eff}\approx  1/(\lambda_1 L)$. The conductance has then the form
\begin{equation}
g\approx N_{eff}\approx \frac{M}{L}[a(\epsilon-\epsilon_{cr})^{\nu_{3d}}M+b]\,.
\end{equation}
This shows that the conductivity in 3D is indeed $\sigma_{xx}\equiv gL/M^2 \sim (\epsilon-\epsilon_{cr})^{\nu_{3d}}$. On the critical line $\epsilon=\epsilon_{cr}$ the conductance is limited to the surface area and is $\sim b$.

Consider next the $\Delta\neq 0$ case with broken spin rotation
symmetry. At $\Delta=2$, e.g., the 2D system ($t=0$) is critical at
$\epsilon_{cr} =\pm 0.64$ with a critical exponent
$\nu_{QH}\approx 2.5$ of the usual quantum Hall system. At $t\neq 0$, we expect
each of the critical states to split into two with a band of
metallic states between them (as for $\Delta =0$), however it is
not obvious wether the two internal critical curves merge or produce a
new phase boundary. We find merging of these lines, producing a
four-phase diagram as shown in Fig. 3. Both outer critical lines
scale as $t\sim |\epsilon - \epsilon_{cr}|^{\nu_{QH}}$ in
agreement with the argument above. The inner curve is affected by
both critical points and therefore deviates from this scaling
form.

Fig. 3 shows the existence of a new phase with $\sigma_{xy}=1$
which becomes metallic at very low values of $t$, e.g. $t=0.001$ at
$\epsilon=0$. This feature can be traced to the rather large
$\xi_M/M$ values for $t=0$ in the range
$-\epsilon_{cr}<\epsilon<\epsilon_{cr}$. We note that at $t=0$ a
single spin state becomes extended at $\epsilon_{cr}$, while the
other spin state becomes extended independently at $-\epsilon_{cr}$.
For $t<0.001$ these extended states produce two metallic bands
which do not overlap, hence when the chemical potential is in
between these bands the $\sigma_{xy}=1$ phase emerges. When
$t>0.001$ these bands overlap and a $\sigma_{xy}=1$ phase is not
possible. We emphasize that there is a single metallic phase,
hence in this phase the two extended spin states mix via 
interlayer coupling, unlike the situation at $t=0$.

\begin{figure}
%[htb]
\begin{center}
\includegraphics[scale=0.7]{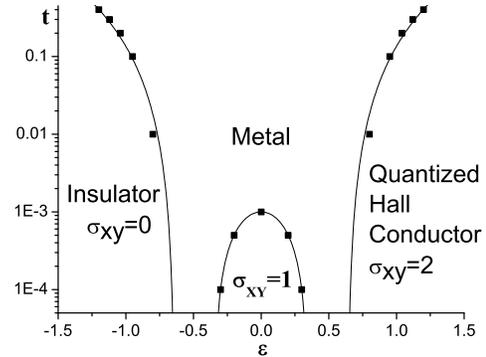}
\end{center}
\caption{Phase diagram of the 3D network model with $\Delta =2$
showing an additional phase with $\sigma_{xy}=1$.}
\end{figure}

We proceed to evaluate the localization
length  exponent 
for $\Delta =2$ and different $t$. For $t=1/\sqrt{2}$
(maximal mixing of Eq. \ref{third}) we find $\nu=0.85$ while
for $t=0.1$ we find $\nu\approx 0.93$, fairly close to 
$\nu_{3d}$. It differs significantly from the value $1.45$ found
for the 3D  $U(1)$ system\cite{cd}. This is consistent with our
finding of a single metallic phase in which extended states with
both spins are mixed. Thus at $t=0$ (extended states at 
$\pm \epsilon_{cr}$ are decoupled) the symmetry is reduced to
$U(1)$ producing the $\nu_{QH}$ exponent, while the $t\neq 0$
metallic phase (mixed extended states) has the same symmetry for $\Delta=0$ or $\Delta\neq 0$.

Based on these results we propose a 3-parameter scaling
function near the multicritical point $\epsilon = \Delta =t=0$ 
\begin{equation}
\frac{\xi_M}{M}=f(\epsilon^{\nu_{2d}}M,\Delta^{\mu}M,t^{\alpha}M)
\end{equation}
On the critical surface the scaling function is $M$ independednt, hence for $\Delta=0$ the critical line is $\epsilon_{cr}\sim t^{\alpha /\nu_{2d}}$. As discussed above, $W(t)\sim\epsilon_{cr}\sim t^{1 /\nu_{2d}}$, therefore $\alpha=1$.

We demonstrate in Fig. 4 scaling along the line $\epsilon = \Delta =0$ with $\alpha=1$. Note that when $t$ is too large, approaching the symmetric point $1/\sqrt{2}$, scaling is not expected. The analysis above yields $\sigma_{xx}\sim t$ when approaching the multicritical point.

Finally we consider a realization of the parameter which drives the phase transitions. Correspondence with edge states of the $d+id'$ system shows that the
transition is driven by the average value $T_0$ of quasiparticle transmission
between grains, which determines $\epsilon$ via $T_0=[1+\exp (-\pi
 \epsilon)]^{-1}$. Following the suggestion that, at least for weakly broken time reversal symmetry, disorder may drive the transitions \cite{senthil1},
we have performed further simulations of the 2D
network searching for the effect of randomness in $\epsilon$. We
have found this randomness to have a negative scaling exponent, i.e. an
irrelevant variable. In fact, the mapping to a percolation
problem\cite{beamond} leads to a distribution in the percolation
parameter which is indeed irrelevant. In our system where time
reversal symmetry breaking is fully developed we do not expect disorder to
drive a spin quantum Hall transition. In contrast, charge transport 
 is dominated by the randomness of the Josephson coupling between superconducting grains;  the phase correlation
 between grains is lost in 2D at a critical value of disorder, as
 shown in an XY model with random phase shifts\cite{carpentier}. 
We conclude then that
a superconductor-insulator transition for charge transport is
disconnected from that of quasiparticle spin transport, realizing
spin-charge separation in transport.

\begin{figure}
%[htb]
\begin{center}
\includegraphics[scale=0.7]{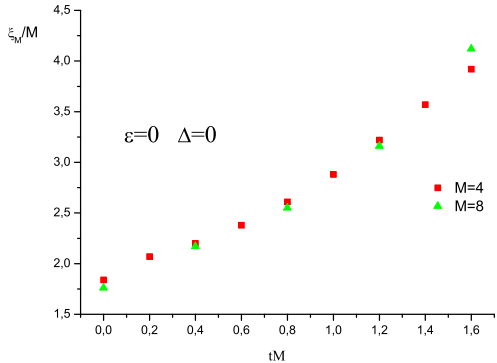}
\end{center}
\caption{Scaling of renormalized localization length for $\epsilon
=\Delta =0$ as function of interlayer coupling.}
\end{figure}

As a concrete realization we consider two grains of $d+id'$
superconductors with parallel edge states along an axis $x$. An
impurity provides an intergrain scattering potential
$Va\delta(x)$, where $a$ is a lattice constant. The right and left
moving edge state $\psi_R(x)$ and $\psi_L(x)$ then satisfy
\begin{eqnarray}\label{edge}
-iv\partial_x \psi_R(x)+Va\delta(x)\psi_L(x)&=&E\psi_R(x)\nonumber\\
 iv\partial_x \psi_L(x)+Va\delta(x)\psi_R(x)&=&E\psi_L(x)
\end{eqnarray}
where $E$ is an energy eigenvalue. Here $v$ is the edge state
velocity, $v\approx a\Delta'$ for a (110) surface and $v\approx
a\Delta$ for a (100) surface\cite{hg} where $\Delta$ and $\Delta'$
are the gaps of $d_{x^2-y^2}$ and $d_{xy}$, respectively, with
$\Delta'\ll \Delta$.

The transmission from an incoming $\psi_R(x)$ to an outgoing
$\psi_L(x)$ is readily evaluated as
\begin{equation}\label{T0}
T_0=\frac{4(Va/2v)^2}{[1+(Va/2v)^2]^2}\,.
\end{equation}
Note that $T_0$ has a maximum of $1$ at $Va/2v=1$ and decreases at
large $V$ [since then the matching of states near the impurity
($E\approx \pm V$) with the nearest levels on the edges ($E\approx
\pm v/a$) is reduced]. The transmission needed for exhibiting an
extended state, $T_0=1/2$, is achieved at $V\approx \Delta'$ for
(110) edges or $V\approx \Delta$ for (100) edges, i.e. a much
weaker coupling in the former case.

In conclusion, we have demonstrated spin-charge separation in transport: spin transport and related QH transitions are controlled by the average intergrain transmission while charge transport and superconducting correlation is controlled by
 the amount of disorder in the intergrain (Josephson) coupling. We show that
interlayer coupling leads to a new spin metal phase and 
identify the localization exponent for the spin insulator-metal
transition as $\nu_{3d}\approx 0.9$. The latter is also the spin conductivity exponent when approaching the transition from the metallic side.

\vspace{2mm}

One of us (V. K.) appreciates valuable discussions with H. Aoki, Y. Hatsugai, T. Nakayama, K. Yakubo and B. Shklovsky. We thank J. L. Pichard for useful comments. Part of this work was done with a support of "FY2003 JSPS Invitation Fellowship Program for Research in Japan (Short-Term)" (V.K.). This research
was supported by THE ISRAEL SCIENCE FOUNDATION founded by the
Israel Academy of Sciences and Humanities.

\end{document}